# Theoretical Studies of Long Lived Plasma Structures

Maxim Dvornikov*†

*Departamento de Física y Centro-Científico-Tecnológico de Valparaíso,
Universidad Técnica Federico Santa María,
Casilla 110-V, Valparaíso, Chile;
†N. V. Pushkov Institute of Terrestrial Magnetism, Ionosphere and
Radowave Propagation (IZMIRAN),
142190, Troitsk, Moscow region, Russia
E-mail: maxim.dvornikov@usm.cl

**Abstract**

We construct the model of a long lived plasma structure based on spherically symmetric oscillations of electrons in plasma. Oscillations of electrons are studied in frames of both classical and quantum approaches. We obtain the density profile of electrons and the dispersion relations for these oscillations. The differences between classical and quantum approaches are discussed. Then we study the interaction between electrons participating in spherically symmetric oscillations. We find that this interaction can be attractive and electrons can form bound states. The applications of the obtained results to the theory of natural plasmoids are considered.

## 1. Introduction

The construction of a theoretical model for a stable spherically symmetric plasma structure is still a puzzle for the plasma physics [1]. The attempts to build a model of a spherical plasmoid with help of the classical Boltzmann and Gibbs statistics result in the formidable difficulties [2]. Moreover the static cluster of charged particles maintained by its own electromagnetic forces is unlikely to be stable since plasma has a tendency to expand at the absence of any external pressure or additional attractive forces like gravity etc. (see Ref. [3]).

We can imagine a spherical plasma structure in the form of a spherically symmetric Langmuir wave [4]. Such a model of a spherical plasmoid has many advantages compared to the static electric charge distribution models. First our model does not require any additional external forces to provide the stability of the system. Second, a charged particles system making the spherically symmetric motion does not loose energy for radiation. Thus, such a spherical plasmoid will be dynamically stable compared to any other unstructured plasma formations since the latter will loose their energy and recombine with the time scale of several milliseconds. Third, in frames of our model one can point out an internal energy source which would compensate the inevitable energy losses and provide the long life-time of a plasmoid.

In this paper we summarize our previous works on the theory of spherically symmetric plasma structures. In Sec. 2 we study spherically symmetric oscillations of electron gas in plasma in frames of both classical and quantum approaches. We obtain the expression for the density of electrons in the explicit form and the dispersion relations for these oscillations. The discrepancy between classical and quantum cases are discussed. In Sec. 3 we discuss the possible formation of a bound state of electrons participating in spherically symmetric oscillations due to the exchange of ion acoustic waves. Finally in Sec. 4 we discuss our results.

## 2. A model of a spherical plasmoid based on electron gas oscillations in plasma

In this section we will show that one can describe spherically symmetric oscillations of electrons in plasma within both classical and quantum approaches. We will obtain the time dependent density profile of electrons and the dispersion relation for these oscillations.

First we consider spherically symmetric oscillations in frames of the classical electrodynamics. We can suggest that only electrons participate in oscillations of plasma since the mobility of ions is low. At the absence of collisions and other forms of dissipation the system of the hydrodynamic equations for the description of the evolution of the electrons velocity $\mathbf{v}$, the electric field $\mathbf{E}$ and plasma pressure $p$ can be presented in the following way [5]:

$$\frac{\partial n_e}{\partial t} + \nabla \cdot (n_e \mathbf{v}) = 0,$$

$$\frac{\partial \mathbf{v}}{\partial t} + (\mathbf{v} \cdot \nabla)\mathbf{v} = -\frac{e}{m}\mathbf{E} - \frac{1}{mn_e}\nabla p,$$

$$\nabla \cdot \mathbf{E} = -4\pi e(n_e - n_i) + 4\pi \rho_{ext}(\mathbf{r},t), \qquad (1)$$

where $n_e$ is the number density of electrons, $n_i = n_0$ is the constant number density of ions, $m$ is the mass on an electron, and $e > 0$ is the proton charge. In Eq. (1) we include the possible external source $\rho_{ext}(\mathbf{r},t)$ as well as take into account that a spherically symmetric system does not have a magnetic field (see Ref. [6]).

If we consider small perturbations of the electrons density $n_e - n_0 = n \ll n_0$ and linearize Eq. (1), we get a single differential equation for the electrons density perturbation [6],

$$\frac{\partial^2 n}{\partial t^2} + \omega_e^2 n - \frac{1}{m}\left(\frac{\partial p}{\partial n}\right)_0 \nabla^2 n = \frac{4\pi e n_0}{m}\rho_{ext}, \qquad (2)$$

where $\omega_e = \sqrt{4\pi n_0 e^2/m}$ is the plasma frequency for electrons and $(\partial p/\partial n)_0$ is the derivative taken at $n_e = n_0$, which depends on the equation of state of electrons gas in plasma.

At the absence of the external source $\rho_{ext}(\mathbf{r},t)$ the spherically symmetric solution of Eq. (2) has the form, $n(r,t) = A_{cl}\cos(\omega t)\sin(\gamma r)/r$, where $A_{cl}$ is the small constant. The frequency of oscillations $\omega$ is related to the length scale parameter $\gamma$ by the following formula,

$$\omega^2 = \omega_e^2 + \frac{1}{m}\left(\frac{\partial p}{\partial n}\right)_0 \gamma^2, \qquad (3)$$

which shows that free oscillations of plasma can exist only if $\omega \geq \omega_e$. Forced oscillation are possible with any frequencies $\Omega$ if we consider the time dependent external source $\rho_{\text{ext}} \sim \cos\Omega t$.

Now we will describe oscillations of electrons in plasma in frames of the quantum approach. We will use the concept of the wave function normalized on the number density of electrons [7], $|\psi(\mathbf{r},t)|^2 = n_e(\mathbf{r},t)$. At the absence of the exchange effects [8] the Schrödinger equation for the description of $\psi(\mathbf{r},t)$ evolution has the form [4],

$$i\hbar\frac{\partial \psi}{\partial t} = \hat{H}\psi,$$

$$\hat{H} = -\frac{\hbar^2}{2m}\nabla^2 + e\varphi(\mathbf{r},t) + e^2\int d^3\mathbf{r}'\frac{|\psi(\mathbf{r}',t)|^2}{|\mathbf{r}-\mathbf{r}'|}, \qquad (4)$$

where $\varphi(\mathbf{r},t)$ is the potential of an external field. In the case when we study the interaction between electrons and ions this potential has the form, $\varphi(\mathbf{r},t) = -e\int d^3\mathbf{r}' n_i(\mathbf{r}',t)/|\mathbf{r}-\mathbf{r}'|$.

As in the previous case we suppose that ions do not participate in plasma oscillations keeping their density unchanged $n_i = n_0$. We will look for a spherically symmetric solution of Eq. (4) using the perturbation theory, $\psi(r,t) = \psi_0 + \chi(r)e^{-i\omega t}$, where $|\psi_0|^2 = n_0$ is the unperturbed density and $\chi$ is the small perturbation of the wave function.

The linearized Eq. (4) for the function $\chi$ can be presented in the following way:

$$\hbar\omega\chi + \frac{\hbar^2}{2m}\left(\chi'' + \frac{2}{r}\chi'\right) - 8\pi e^2 n_0 \int_r^\infty \frac{dx}{x^2}\int_0^x y^2\chi(y)dy = 0. \qquad (5)$$

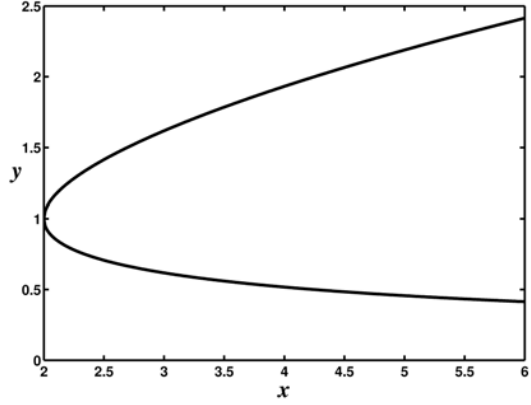

**FIGURE 1**. The coefficient $y = \gamma\sqrt{\hbar/2\omega_e m}$ versus the parameter $x = \omega/\omega_e$.

The solution of Eq. (5) has the form, $\chi(r) = A_q \sin(\gamma r)/r$, where $A_q$ is the small constant. The density of electrons can be calculated on the basis of the obtained wave function, $n_e(r,t) = n_0 + A_q \cos(\omega t)\sin(\gamma r)/r + \cdots$, which coincides with the classical case studied earlier.

The frequency of quantum oscillations $\omega$ depends on the parameter $\gamma$ as

$$\gamma^2 = \frac{\omega m}{\hbar}\left[1 \pm \left(1 - 4\frac{\omega_e^2}{\omega^2}\right)\right]. \tag{6}$$

One can see that the dispersion relation (6) is different from Eq. (3) and has two branches (see Fig. 1). The upper branch corresponding to the sign "+" in Eq. (6) can be treated as the classical one since at high oscillations frequencies one gets $\hbar\omega = (\hbar\gamma)^2/2m$ as for a classical particle. The quantum effects are likely to be responsible for the appearance of the lower branch corresponding to the sign "−" in Eq. (6) since $\gamma \to \sqrt{2m\omega_e^2/\hbar\omega}$ for $\omega \gg \omega_e$, i.e. the Plank constant cannot be eliminated. One can also see on Fig. 1 that within the quantum approach oscillations of electrons can exist if $\omega \geq 2\omega_e$ whereas in classical approach the critical frequency is $\omega = \omega_e$ [see Eq. (3)].

## 3. Interaction between oscillating electrons at the initial stages of the spherical plasmoid evolution

In the previous section we studied steady-state oscillations of electrons in plasma. It was shown that free oscillations with $\omega \sim \omega_e$ are possible. Accounting for the fact that the Langmuir frequency is about 100 GHz for plasma with electrons density $\sim 10^{15}\,\text{cm}^{-3}$, we get that to create a plasmoid one should use a high voltage and very high frequency generator. However spherical plasmoids appear in natural conditions and it is quite difficult to find a natural "generator" with such characteristics. Therefore one should point out a physical process, acting at the stages of a plasmoid formation, which makes the generation of a plasmoid possible.

It was mentioned in Ref. [9] that dense plasma can reveal superconducting properties and the implications of this phenomenon to the natural plasmiods theory were studied there. We can also suggest that at the initial stages of a spherical plasmoid evolution the plasma is in the superconducting stage. The superconductivity will reduce possible friction mechanisms and facilitate the creation of a plasmoid.

It is known that the formation of a bound state of two electrons, a Cooper pair, underlies the superconductivity phenomenon in metals. Usually a Cooper pair is destroyed when the temperature of a metal exceeds a few Kelvin degrees. Although the temperature of plasmas in natural conditions is far greater than a typical temperature of a superconducting metal one can find physical processes leading to the appearance of the effective attraction between electrons. A charged test particle, e.g., an electron, moving in plasma is known to emit ion acoustic waves. Therefore a test electron can be surrounded by a cloud of positively charged ions. Under certain conditions this effective potential can screen the repulsive interaction between two electrons and result in the creation of a bound state (see Ref. [10]). This phenomenon is analogous to the Cooper pairs formation [11].

Let us study the electric field potential created by electrons participating in spherically symmetric oscillations considering each electron as a test particle with the charge $q$. We suppose that a test particle makes forced harmonic oscillations around the point $\mathbf{r}_0$ with the frequency $\Omega$ and the amplitude $\mathbf{a}$: $\mathbf{r}(t) = \mathbf{r}_0 + \mathbf{a}\sin\Omega t$. Each of the test particles is taken to interact with the rest of hot electrons, having the temperature $T$, and with cold ions. The permittivity of this plasma has the form [10],

$$\varepsilon(\mathbf{k},\omega) = 1 + \left(\frac{k_e}{k}\right)^2 - \left(\frac{\omega_i}{\omega+i0}\right)^2, \qquad (7)$$

where $k_e = \sqrt{4\pi n_0 e^2/T}$ is the Debye wave number and $\omega_i = \sqrt{4\pi(Ze)^2 n_0/M}$ is the plasma frequency for ions with the mass $M$ and the charge $Ze$, $Z$ is the degree of the ionization of an ion.

Let us choose the specific coordinate system with $\mathbf{r}_0 = 0$ and $\mathbf{a} = a\mathbf{e}_z$, where $\mathbf{e}_z$ is the unit vector along the $z$ axis. The scalar potential of the electric field of such a system can be presented as a sum of two terms, $\varphi = \varphi_D + \varphi_W$. It is convenient to express these functions in the cylindrical coordinates $\mathbf{r} = (\rho, z, \phi)$.

The function $\varphi_D$ is the analog of the Debye-Hückel potential [6],

$$\varphi_D(\rho,z,t) = 2q\sum_{n=0}^{\infty}(-1)^n\int_0^{\infty}k_\rho dk_\rho \frac{e^{-|z|\sqrt{k_\rho^2+k_e^2}}J_0(k_\rho\rho)}{\sqrt{k_\rho^2+k_e^2}}$$
$$\times\left(I_{2n}\left(a\sqrt{k_\rho^2+k_e^2}\right)\cos[2n\Omega t]\right.$$
$$\left.\pm I_{2n+1}\left(a\sqrt{k_\rho^2+k_e^2}\right)\sin[(2n+1)\Omega t]\right), \qquad (8)$$

where $J_0(x)$ is the Bessel of zero order and $I_n(x)$ are the Bessel functions of the imaginary argument. In Eq. (8) the sign "+" stands for $z>0$ and "–" – for $z<0$. Note that due to the axial symmetry this potential does not depend on the angular coordinate $\phi$. It is possible to show that in the static case, when $a\to 0$, the function $\varphi_D$ takes the form, $\varphi_D = (q/r)\exp(-k_e r)$, where $r=\sqrt{\rho^2+z^2}$, i.e. it reproduces the well known Debye-Hückel potential.

The wake potential $\varphi_W$ can be represented in the cylindrical coordinates as (see Ref. [6])

$$\varphi_W(\rho,z,t) = -2q\int_0^{\infty}k_\rho dk_\rho \frac{e^{-|z|\sqrt{k_\rho^2+k_e^2}}J_0(k_\rho\rho)}{\sqrt{k_\rho^2+k_e^2}}I_0\left(a\sqrt{k_\rho^2+k_e^2}\right)$$
$$+2q\sum_{n=1}^{\infty}\begin{Bmatrix}1\\(-1)^n\end{Bmatrix}\frac{\omega_i^2}{\omega_i^2-(n\Omega)^2}\frac{k_n^3}{k_e^2+k_n^2}$$
$$\times\int_0^1 dx J_0\left(k_n\rho\sqrt{1-x^2}\right)J_n(ak_n x)\sin(k_n|z|x-n\Omega t), \qquad (9)$$

where the upper multiplier corresponds to $z>0$, the lower one – to $z<0$, $J_n(x)$ are the Bessel functions and $k_n = k_e n\Omega/\sqrt{\omega_i^2-(n\Omega)^2}$. To derive Eq. (9) we suppose that $n\Omega<\omega_i$.

It should be noted that the potentials (8) and (9) are different from the analogous expressions found in Ref. [12] where the radiation of a charged linear oscillator in vacuum was studied.

This discrepancy is because of the distinct gauges. It was shown in Ref. [6] that for the system making spherically symmetric motion it is convenient to choose the gauge $\mathbf{A}=0$, which is adopted here, whereas in Ref. [12] the Lorentz gauge was used.

Comparing Eqs. (8) and (9) we see that the time independent terms cancel there. Let us suppose that we excited only the first harmonic of ion acoustic waves with $n=1$. The wake potential at $\rho=0$, i.e. at the line of test particle oscillations, takes the form,

$$\varphi_W(z,t) = \mp q \frac{\Omega^2}{\omega_i^2 - \Omega^2} \frac{k_1 a}{|z|} \cos(k_1 |z| - \Omega t), \quad (10)$$

where the sign "−" stands for $z>0$ and "+" – for $z<0$. The corresponding contribution to the Debye-Hückel potential in Eq. (8) is negligible at the distances $r > 1/k_e$.

Now we can see that the potential (10) can be attractive if $\cos(k_1|z|-\Omega t)>0$ for $z>0$ and if $\cos(k_1|z|-\Omega t)<0$ for $z<0$. For a test electron ($q=-e$) to form a bound state with another electron in plasma the wake potential should be greater than the kinetic energy of its oscillations, $|e\varphi_W| > E_{osc} = ma^2\Omega^2/2$. Suppose that two electrons are at the distance $d = 1/k_e$ and the amplitude of oscillations $a = 0.1d \ll d$. Studying the plasma consisting of electrons, with the temperature $T = 10^3$ K and the number density $n_0 = 10^{15}$ cm$^{-3}$, as well as of singly ionized nitrogen atoms we get that a bound state can be formed if $10^6\,\mathrm{s}^{-1} < \Omega < 10^{10}\,\mathrm{s}^{-1}$. Thus we get that the effective attraction can take place in the atmospheric plasma for the reasonable frequencies of an external field.

## 4. Discussion

In this work we have summarized our recent theoretical research on spherically symmetric long lived plasma structures. In

Sec. 2 we developed the model of a plasmoid based on spherically symmetric oscillations of electrons in plasma. We studied stable oscillations of electrons in frames of both classical and quantum approaches. It was obtained that the behavior of oscillations is similar in classical and quantum cases, however the dispersion relations (3) and (6) are different. This discrepancy is due to the appearance of quantum effects in the lower branch of the dispersion relation (6).

If we study the second correction to the number density of electron within the quantum approach, one gets that the static uncompensated negative volume charge appears in the center of the system [4], $\langle n_e(r,t) \rangle = n_0 + |\chi(r)|^2$. This excessive negative charge will be compensated by the volume charge of positive ions, i.e. the static density of ions will increase in the center of the system. The analysis presented in Sec. 2 is based on the approximate solution of the linearized Schrödinger equation. It means that it just shows the tendency of the ions density enhancement.

In Ref. [4] we put forward a hypothesis that in realistic situation the central density of ions of a spherical plasmoid can be big enough to initiate micro-dose nuclear fusion reactions. These reactions can serve as an internal energy source of a plasmoid. The idea that nuclear reactions can provide an energy supply for natural spherical plasma structures was discussed in Ref. [13].

It was found in Ref. [14] that a monochromatic Langmuir wave is unstable and the density of electrons can achieve huge values during a finite time. Of course, in realistic situation the infinite values of electrons density are impossible. During the collapse of Langmuir waves their energy is transferred to electrons providing the plasma heating due to the various dissipation mechanisms like Landau damping. This phenomenon might be important at some stages of the spherical plasma structures evolution.

In Sec. 3 we discuss the interaction between electrons participating in forced spherically symmetric oscillations through the exchange of ion acoustic waves. It was found that under certain conditions the exchange of ion acoustic waves can screen the repulsive Debye-Hückel potential and result in the effective attraction between two electrons. The attractive potential can lead to the formation of a bound state of two electrons, an analog of a Cooper pair, and possible superconducting phase inside a spherical plasmoid. The existence of the superconducting phase inside natural plasmoids was previously proposed in Ref. [9]. However in these papers the dense plasma was already supposed to be in the superconducting state and the phenomenological consequences of this phenomenon were considered. In our work we point out the physical mechanism which can underlie the plasma superconductivity phenomenon.

## Acknowledgements

The work has been supported by the Conicyt (Chile), Programa Bicentenario PSD-91-2006. The author is thankful to the organizers of AIS-2010 conference for the invitation.